\documentclass[conference]{IEEEtran}

\newif\ifanonymous
\anonymousfalse 

\usepackage[inline]{enumitem}
\usepackage{xspace}
\usepackage{tikz}
\usetikzlibrary{tikzmark,arrows.meta,calc}
\usepackage{array}              
\usepackage{pifont}             
\usepackage{url}
\usepackage{orcidlink}
\usepackage[T1]{fontenc} 
\usepackage{amsmath}%
\usepackage{wasysym}%
\usepackage[capitalise,nameinlink,noabbrev]{cleveref}%
\usepackage{booktabs}
\usepackage{boxedminipage}

\usepackage{colortbl}
\usepackage{amssymb}
\usepackage{inconsolata} 

\usepackage{hyperref}
\usepackage{flushend}

\usepackage{relsize}  
\def\EFDD{{\smaller{}EFDD}\xspace}
\def\SFLKIT{{SFLKit}\xspace}
\def\TESTS4PY{{Tests4Py}\xspace}
\def\TARANTULA{{\smaller{}TARANTULA}\xspace}
\def\OCHIAI{{\smaller{}OCHIAI}\xspace}
\def\DSTAR{{\smaller{}DSTAR}\xspace}
\def\NAISHTWO{{\smaller{}NAISH2}\xspace}
\def\GPTHIRTEEN{{\smaller{}GP13}\xspace}
\def\EXAM{{\smaller{}EXAM}\xspace}

\definecolor{Grey}{rgb}{0.5,0.5,0.5}
\definecolor{LightGrey}{rgb}{0.9,0.9,0.9}
\definecolor{Green}{rgb}{0.0,0.6,0.0}
\definecolor{Red}{rgb}{0.6,0.0,0.0}
\definecolor{Blue}{rgb}{0.0,0.0,0.6}

\definecolor{DarkBlue}{rgb}{0.0859, 0.308, 0.523}
\definecolor{DarkOrange}{rgb}{0.8, 0.4, 0.0}
\definecolor{DarkGreen}{rgb}{0.00,0.40,0.00}
\definecolor{ScarletRed}{rgb}{0.60,0.00,0.00}
\definecolor{AlmostWhite}{rgb}{0.80,0.80,0.80}
\definecolor{Gray}{gray}{0.85}

\definecolor{row}{RGB}{252, 235, 207} 

\newcommand{\PASS}{{\color{Green}\ding{52}}\xspace}
\newcommand{\FAIL}{{\color{Red}\ding{56}}\xspace}

{\medskip
\noindent
\let\emph=\textbf
\begin{boxedminipage}{\linewidth}\begin{center}\em}%
{\end{center}\end{boxedminipage}%
\medskip
}

\newlist{questions}{enumerate}{1}
\setlist[questions,1]{label=\bfseries RQ\arabic*:,ref=RQ\arabic*,leftmargin=3\parindent}
\crefname{question}{}{}
\Crefname{question}{}{}

\makeatletter
\newcommand{\linebreakand}{%
    \end{@IEEEauthorhalign}
    \hfill\mbox{}\par
    \mbox{}\hfill\begin{@IEEEauthorhalign}
}
\makeatother

\title{How do Execution Features Improve Statistical Fault Localization?\\\LARGE An Empirical Study}

\ifanonymous
\author{
    \IEEEauthorblockN{Anonymous Author(s)}
}
\else
\author{
    \IEEEauthorblockN{Marius Smytzek \orcidlink{0000-0002-4899-9031}}
    \IEEEauthorblockA{CISPA Helmholtz Center for Information Security\\
 Saarbr{\"u}cken, Germany\\
 marius.smytzek@cispa.de
 }
    \and
    \IEEEauthorblockN{Andreas Zeller \orcidlink{0000-0003-4719-8803}}
    \IEEEauthorblockA{CISPA Helmholtz Center for Information Security\\
 Saarbr{\"u}cken, Germany\\
 andreas.zeller@cispa.de
 }
}
\fi

\begin{document}
    \maketitle

    \begin{abstract}
Automated fault localization helps developers find faults in large code bases.
Statistical fault localization (SFL) ranks suspicious lines from pass/fail spectra, but line execution alone misses information like data-flow, values, or branch conditions that explain why a failure occurs.

This study evaluates whether augmenting SFL with execution features improves localization accuracy and developer-oriented inspection effort.
We extract execution features with \EFDD{} for all \TESTS4PY{} subjects, train per-subject random forests, map importances to source lines, and combine the resulting weights with established SFL formulas.
The evaluation measures reference-patch accuracy, line- and function-level effort, robustness, and feasibility using a confounder-adjusted mixed-effects model, corroborated by paired statistical tests and outcome-neutral quality checks.
    \end{abstract}

    \begin{IEEEkeywords}
 Fault localization, automated debugging, execution features, diagnosis, dynamic analysis
    \end{IEEEkeywords}

    \section{Introduction}%
    \label{sec:intro}

 Debugging consumes a significant portion of development effort, requiring developers to identify root causes across thousands of lines of code.
 Automated fault localization techniques address this by ranking suspicious code locations most likely responsible for failures.
 Traditional statistical fault localization (SFL) techniques, such as \TARANTULA{}~\cite{jones2002tarantula}, correlate code line execution with test outcomes and have become widely adopted because code locations are directly actionable for developers.
 However, SFL cannot distinguish among many lines executed in failing tests, passing tests weaken correlations for shared lines, and ranked lists do not explain failure conditions.
 Our prior work on Execution-Feature-Driven-Debugging (\EFDD) empirically demonstrated that execution characteristics beyond line coverage, including definition-use pairs~\cite{santelices2009defuse}, variable value predicates, and scalar-pair relations~\cite{liblit2005sd}, correlate more strongly with failures than line execution alone.

\begin{figure}
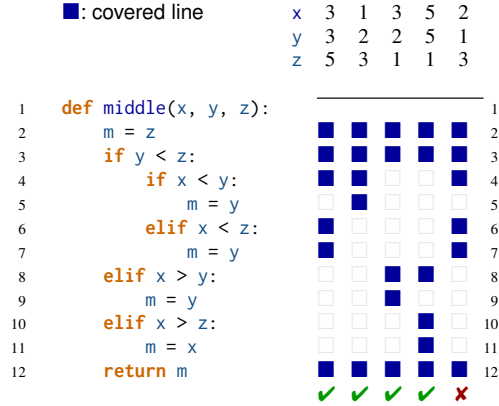

        \def\*{{\color{Blue}$\blacksquare$}}
        \def\+{{\color{Blue}$\blacksquare$}}
        \def\-{{\color{LightGrey}$\Box$}}
        \def\ind{\qquad}
        \scriptsize
        \centering
        \resizebox{.777\columnwidth}{!}{
        \begin{tabular}{@{}>{\tiny}r>{\tt}l@{\quad}l@{\ \ }r@{\ \ }r@{\ \ }r@{\ \ }r@{\ \ }r@{\ \ }>{\tiny}r}
            \\
            & \textsf{\color{black}\*: covered line}
            & \color{Blue}\texttt{x}      & 3   & 1   & 3   & 5   & 2  \\
            & & \color{DarkBlue}\texttt{y}  & 3   & 2   & 2   & 5   & 1  \\
            & & \color{DarkBlue}\texttt{z} & 5   & 3   & 1   & 1   & 3  \\
            \\ \cline{4-8}
 1  & \textbf{\color{DarkOrange}def} {\color{Blue}middle}({\color{DarkBlue}x}, {\color{DarkBlue}y}, {\color{DarkBlue}z}):  & & & & & & & 1 \\
 2  & \ind {\color{DarkBlue}m} = {\color{DarkBlue}z}
            & & \*  & \*  & \*  & \*  & \*  & 2 \\
 3  & \ind  \textbf{\color{DarkOrange}if} {\color{DarkBlue}y} < {\color{DarkBlue}z}:
            & & \*  & \*  & \*  & \*  & \*  & 3 \\
 4  & \ind \ind  \textbf{\color{DarkOrange}if} {\color{DarkBlue}x} < {\color{DarkBlue}y}:
            & & \*  & \*  & \-  & \-  & \*  & 4 \\
 5  & \ind \ind \ind {\color{DarkBlue}m} = {\color{DarkBlue}y}
            & & \-  & \*  & \-  & \- & \-  & 5 \\
 6  & \ind \ind  \textbf{\color{DarkOrange}elif} {\color{DarkBlue}x} < {\color{DarkBlue}z}:
            & & \*  & \-  & \-  & \-  & \*  & 6 \\
 7  & \ind \ind \ind {\color{DarkBlue}m} = {\color{DarkBlue}y}
            & & \*  & \-  & \-  & \-  & \*  & 7 \\
 8 & \ind  \textbf{\color{DarkOrange}elif} {\color{DarkBlue}x} > {\color{DarkBlue}y}:
            & & \-  & \-  & \*  & \*  & \-  & 8 \\
 9 & \ind \ind {\color{DarkBlue}m} = {\color{DarkBlue}y}
            & & \-  & \-  & \*  & \-  & \-  & 9 \\
 10 & \ind  \textbf{\color{DarkOrange}elif} {\color{DarkBlue}x} > {\color{DarkBlue}z}:
            & & \-  & \-  & \-  & \*   & \-  & 10 \\
 11 & \ind \ind {\color{DarkBlue}m} = {\color{DarkBlue}x}
            & & \-  & \-  & \-  & \*  & \-  & 11 \\
 12 & \ind  \textbf{\color{DarkOrange}return} {\color{DarkBlue}m}
            & & \*  & \*  & \*  & \*  & \*  & 12 \\
            &
            & & \PASS{} & \PASS{} & \PASS{} & \PASS{} & \FAIL{}\\
        \end{tabular}
 }%
        \caption{Fault Localization.
 SFL ranks Lines~6 (correct) and 7 (fault) equally high while missing the condition \texttt{x} > \texttt{y} provided by execution features.}%
        \label{fig:middle-example}
    \end{figure}

 Consider the \texttt{middle()} function (\Cref{fig:middle-example}), which returns the median of its inputs.
 When \texttt{middle(2, 1, 3)} fails (returning 1, not 2), its executed lines also appear in passing runs, so SFL ranks Lines 6 and 7 as most suspicious but cannot explain why.
 Execution features reveal the critical difference: reaching Line~7 while \texttt{x} > \texttt{y} holds occurs only in failing tests, exposing the condition that coverage alone misses.

 We propose augmenting SFL with execution feature-based weighting.
 Our method extracts execution features from passing and failing test runs, trains a classifier to identify failure-indicative features, maps features to source locations, and weights baseline SFL scores by feature importance.
 The baseline formula remains intact, so the method adds an inspectable weighting signal rather than replacing SFL.
 It contributes a pre-registered evaluation of whether feature augmentation transfers \EFDD's diagnostic evidence to localization, spanning line- and function-level accuracy, separate reference-patch and test-impact outcomes, and practical utility through diagnostics and quality checks.

\section{Background and Related Work}%
\label{sec:background-related-work}

\subsection{Statistical Fault Localization}

SFL ranks program elements by correlating lines with test outcomes.
Classical formulas include Tarantula~\cite{jones2005tarantula}, Ochiai~\cite{abreu2006ochiai}, DStar~\cite{wong2012dstar}, and Naish2~\cite{naish2011sbfl}, with early evaluations analyzing formula accuracy and inspection effort~\cite{abreu2007accuracy}.
\SFLKIT{} provides a configurable Python workbench with multiple predicate types, spectra, and formulas~\cite{smytzek2022sflkit}.
Comparative studies show that relative performance depends on datasets, fault types, and evaluation protocol, with real faults differing from older artificial settings~\cite{pearson2017sfl}, which motivates evaluating feature augmentation across several formulas rather than treating one suspiciousness coefficient as representative.

\subsection{Learning and Reweighting for Fault Localization}

Prior work improves SFL by reweighting spectra, combining spectra with search or optimization, or learning over program elements.
Examples include PageRank-based test differentiation~\cite{zhang2017pagerank}, genetic programming and search-based localization~\cite{xie2013gp}, constrained feature selection over traces~\cite{le2015topk}, and learned rankings from dynamic, mutation-based, and static features~\cite{kim2019prince}.
Our study follows this additive view but differs from prior learned-ranking methods on a critical point.
Prior methods learn the ranking itself and effectively replace the suspiciousness formula with a learned model.
We keep each baseline formula and only modulate it multiplicatively, use the random forest only for per-subject feature attribution, and keep the added signal inspectable and separable from the baseline.

\subsection{Execution Features and Data-Flow Signals}

Execution features extend line coverage with various dynamic information.
For instance, def-use information can strengthen localization by capturing dependencies missed by coverage-only spectra~\cite{santelices2009defuse}.
Predicate-based localization, such as PredFL, shows that branch conditions complement spectrum-based localization~\cite{jiang2019combining}, and TraPT combines program/test transformations with learning-to-rank~\cite{li2017transforming}.
Our prior \EFDD work showed that various execution features capture failure conditions that line spectra miss~\cite{smytzek2025efdd}, motivating feature weighting that remains compatible with existing SFL pipelines.

\subsection{Complementary Debugging Signals}

Our central claim, that richer execution information improves localization, connects to several established lines.
Statistical debugging isolates bugs by correlating instrumented predicates with failures across runs~\cite{liblit2005sd}, and program slicing narrows attention to statements that dynamically or statically affect a faulty value~\cite{soremekun2021slicing}.
Value-based localization similarly reasons over variable states rather than line coverage alone.
These approaches change either the granularity (predicates, slices, values) or the inference (correlation, dependence). In contrast, we keep the established SFL ranking intact and add an inspectable feature-importance weight on top of it.
Beyond spectra and execution predicates, specification-assisted localization introduces violated specifications~\cite{gopinath2012specifications}, while learning-based repair connects localization with patch generation, ranking, and validation~\cite{zhang2023learning}, reinforcing the need to evaluate localization signals independently before combining them with downstream repair machinery.

\subsection{Evaluation Practice and Practical Utility}

Fault-localization evaluation can overstate improvements through single-fault simplification, idealized oracles, and effort models~\cite{steimann2013threats,soremekun2023evaluating}.
Ranked suspiciousness alone does not always help developers~\cite{parnin2011automated}.
A critical limitation is \textit{single-patch bias}, where evaluating only against reference edits can miss broader failure-relevant regions.
We therefore complement standard top-$k$ and \EXAM{}, kept for comparability, with function-level aggregation, test-impact coverage, diagnostics, and outcome-neutral checks, so conclusions do not depend on one reference patch alone.

\subsection{Positioning of This Study}

Existing studies show that reweighting and learning improve rankings and that execution features provide richer signals than line execution alone, yet two gaps remain: whether feature-importance weighting yields robust gains across multiple SFL baselines on real faults, and whether conclusions hold beyond reference-patch-only evaluation.
This study addresses both for \TESTS4PY{}.
Because a positive result is not guaranteed (mapping is heuristic, importances may be unstable, and sparse failing tests may weaken learning), we separate patch-line, function-level, test-impact, and diagnostic outcomes to distinguish true gains from longer functions, frequent execution, or benchmark-patch idiosyncrasies.
We scope the study to a \emph{controlled paired delta}: execution-feature evidence as an additive signal over established SFL, holding code, tests, formula, and metrics fixed.
LLM-based localization is a complementary future direction~\cite{kang2024llm}, but as a baseline it would confound this delta with prompt design, model priors, and reranking, so measuring the delta is a prerequisite for, not a substitute for, later LLM-integrated localization.

\section{Research Questions and Hypotheses}%
\label{sec:research-questions}

\subsection{General Hypotheses}

\begin{description}
   \item[$H_{\mathrm{ACC}}$ \textbf{(Accuracy Hypothesis).}] Augmenting SFL with execution-feature weights improves fault localization effectiveness over baseline SFL, because execution features capture failure-relevant behavior that line spectra alone cannot represent.
   \item[$H_{\mathrm{ROB}}$ \textbf{(Robustness Hypothesis).}] The improvement is robust across different SFL formulas, because execution-feature signals are orthogonal to the suspiciousness formula used by the baseline.
\end{description}

\subsection{Research Questions}

\begin{questions}
   \item \label[question]{rq1}
   \textbf{Patch-Line Accuracy.}
Does execution-feature augmentation improve top-$k$ localization metrics (top-1, top-5, top-10) for reference-patch lines in \TESTS4PY{} compared with each baseline SFL technique?

   \item \label[question]{rq2}
   \textbf{Developer Effort.}
Does execution-feature augmentation improve or maintain line- and function-level inspection-effort metrics compared with each baseline SFL technique?

   \item \label[question]{rq3}
   \textbf{Broader Ground Truth.}
Do augmented rankings discover failure-relevant locations beyond the reference patch, as measured by dynamically discovered test-impact sets?

   \item \label[question]{rq4}
   \textbf{Robustness.}
Is the improvement direction consistent across all five baseline techniques (\TARANTULA{}, \OCHIAI{}, \DSTAR{}, \NAISHTWO{}, and \GPTHIRTEEN{}) and, exploratorily, across the implemented ranking variants?

\end{questions}

\section{Variables, Datasets, and Sampling}%
\label{sec:variables}

\subsection{Independent Variables}

We vary two factors: localization method and baseline formula (\TARANTULA{}, \OCHIAI{}, \DSTAR{}, \NAISHTWO{}, and \GPTHIRTEEN{}).
The localization methods are exactly the unmodified SFL baselines and the primary random-forest feature weighting, plus a clearly separated, exploratory family of ranking variants defined in \Cref{sec:diagnosis-based-fault-localization}.

\subsection{Dependent Variables}

\textit{Patch-line outcomes} (\Cref{rq1} for top-$k$, \Cref{rq2} for effort) are top-1, top-5, and top-10 accuracy, \EXAM{} score, and wasted effort (best, average, worst case) against reference-patch lines.
\textit{Function-level outcomes} (\Cref{rq2}) aggregate rankings at function granularity to reflect developer workflows: function-weighted mass@1/3/5, faulty-function probability mass, function-weighted inspection effort, and practical effort until reaching a faulty function.
\textit{Test-impact outcomes} (\Cref{rq3}) evaluate against both reference patch hunks and dynamically discovered test-impact sets: impact-found@1/5/10, minimum impact rank, and the total number of impact locations ranked.
\textit{Exploratory ranking-variant outcomes} (\Cref{rq4}, exploratory) apply the same patch-line and test-impact metrics to the ranking variants of \Cref{sec:diagnosis-based-fault-localization}.
\textit{Diagnostic metrics} are the interpretive quantities enumerated in \Cref{sec:analysis-plan} (importance concentration, entropy, skewness, dropping, mapping conflicts, and weak-signal counts).
We additionally report feasibility outcomes for runtime, timeouts, exclusions, and trace/storage failures.
Ties in ranking are handled through repeated randomized tie-breaking and averaged per subject.

\subsection{Confounding Variables and Controls}

The confounders we consider are exactly bug difficulty, project size, executable-line count, test-suite size, number of failing tests, coverage density, and the chosen SFL formula.
Our design is within-subject paired: for each bug, the baseline and the augmented method run on identical code, tests, and spectra, so every time-invariant per-bug characteristic in this list takes the same value for both methods and does not shift the average paired contrast.
Pairing does not make these factors irrelevant, however; they govern how large the augmentation effect is and where it concentrates, so we model them rather than set them aside.
Our primary analysis is therefore the confounder-adjusted mixed-effects model of \Cref{sec:analysis-plan}, in which the method indicator is the fixed effect of interest, these confounders enter as covariates, the SFL formula enters as a fixed factor, and random effects per project and subject absorb residual between-cluster variation.
Line- and function-level inspection effort is additionally normalized by the number of executable lines and functions, respectively, so the covariate reflects difficulty rather than raw size.
A per-project random slope for the method effect estimates effect heterogeneity directly, and we report per-project distributions together with the diagnostics in \Cref{sec:variables} to show where sparse failing evidence or feature dropping drives weak or negative effects.
The distribution-free paired tests retain their role as assumption-light corroboration of this model.
We keep these metric families separate, rather than collapsing them into one score, because a method may improve line-level ranking while degrading effort or broader coverage.

\subsection{Dataset, Sampling, and Unit of Analysis}

We evaluate all 310 reproducible \TESTS4PY{} bugs without subsampling because \TESTS4PY{} provides executable Python faults with tests and reference fixes, and using the complete set avoids selection bias.
Each bug is one analysis unit.

\subsection{Operational Definitions}

Reference faulty lines are the exact lines modified in the official benchmark fix. Patch-hunk impact sets contain the corresponding modified hunk regions.
For function-level outcomes, every function containing at least one reference faulty line or patch-hunk line is treated as faulty so that multi-location patches may yield multiple faulty functions.
Test-impact sets use observed line spectra. A line is included if at least one failing test observes it and either no passing test observes it or its failing-observation ratio is at least 0.6.
If this yields an empty set, we use all failing-observed lines.
We choose 0.6 because a line observed in a clear majority of failing tests is more plausibly failure-relevant. 
At the same time, lower thresholds admit incidental lines and higher ones discard relevant lines exercised by only some failing tests.
We vary it over the fixed grid $\{0.5, 0.6, 0.7, 0.8\}$ as a pre-registered sensitivity analysis.
We keep patch-hunk and test-impact sets distinct so that an improvement against reference edits is not automatically read as improved coverage of failure-relevant behavior.

\section{Methodology}%
\label{sec:diagnosis-based-fault-localization}

Our core approach extracts execution features, trains a random forest to identify failure-correlated features, maps feature importances to source locations, and boosts baseline SFL scores with feature-derived weights.

\subsection{Feature Extraction}

A \emph{feature} is an \EFDD{} execution-derived signal encoded as a binary or tertiary variable.
In line with the \EFDD{} feature taxonomy, features are drawn from coverage, value, condition, and exception families, including line/branch/function/loop execution, def-use pairs, variable-value and length properties, scalar-pair relations, condition outcomes, and exception exits~\cite{smytzek2025efdd}.
We use \EFDD{} to instrument Python code and generate one labeled feature vector per passing or failing test execution.
\EFDD{} aggregates an entire run into a single vector recording the presence and values of these features, matching the run's pass/fail label.
One labeled vector per execution is the natural unit at test-outcome granularity, and a passing and a failing vector suffice to compute feature differences, so no within-run subdivision is required.
Unobserved features receive their fixed default value, requiring no distributional assumptions.
Constant within-subject features are excluded before primary random-forest training and reported as feature-dropping diagnostics.

\subsection{Random Forest Model Training}

For each subject, we train a 200-estimator scikit-learn random forest (RF) with Gini splits on all available labeled vectors after constant-feature removal.
We do not tune hyperparameters per subject.
All other random-forest parameters remain at their defaults except balanced class weights to account for pass/fail imbalance when a subject has few failing tests, and random operations use master seed 42 or deterministic seeds derived from it.
We use a random forest because execution features are sparse, correlated, and discriminative only in combination: it captures non-linear interactions and yields importance scores that map back to source locations.

The RF is a per-subject feature-\emph{attribution} device, not a predictor of unseen data: we extract failure-associated importances within one known faulty revision and read them as per-line weights modulating an unchanged SFL formula, so classical overfitting to unseen executions is not the central validity concern.
We therefore train per subject with no train/test split; standard SFL is itself transductive, computing suspiciousness from the same pass/fail spectra, so a held-out split would evaluate the augmentation on a different footing than its baseline.
The design also resists spurious importance: bagging and per-split feature subsampling reduce variance, all features are binary or tertiary so the impurity-importance bias toward high-cardinality features does not apply, and the negative control (\Cref{sec:outcome-neutral-checks}) would expose any gain that is merely a fitting artifact.
The model thus estimates failure-informative weights rather than final ranks, keeping the result an SFL ranking augmented by execution evidence, not a replacement.

\subsection{Feature Importance Extraction and Mapping}

We extract scikit-learn feature importances and map features to lines by semantics, strictly following the mapping of \EFDD{}~\cite{smytzek2025efdd}, for instance, line features map to their own line, def-use pairs map to both definition and use lines, or value predicates map to definition lines.
These mapping rules are fixed before execution and applied uniformly across all subjects and formulas.
If one feature maps to multiple source lines, the same feature importance is assigned to each mapped line before aggregation.
Raw feature importances are min-max normalized to $[\varepsilon,1]$ with $\varepsilon=10^{-3}$ as $w_f=\varepsilon + (1-\varepsilon)(I(f)-I_{\min})/(I_{\max}-I_{\min})$.
If all importances are equal, we use $I_{\min}=0$ and $I_{\max}=1$.
If multiple features map to a line, we take the maximum normalized weight, which prevents many low-value features from inflating a line while still letting one strong failure-indicative feature affect the ranking.
Lines without mapped evidence receive the lower bound $\varepsilon$.
We record mapping conflicts and feature dropping as diagnostics and relate conflict rates to performance changes.

\subsection{Score Boosting}

We first compute $\text{susp}_{w}(l)=\text{susp}'(l)\times w(l)$, where $\text{susp}'(l)$ is the original SFL score and $w(l)$ the feature weight, normalized within each subject and baseline so formula scale does not drive the blend.
The primary RF ranking then uses $\text{susp}(l)=(1-\lambda)\operatorname{norm}(\text{susp}'(l))+\lambda\operatorname{norm}(\text{susp}_{w}(l))$ with a priori $\lambda=0.35$.
Thus $1-\lambda=0.65$ keeps baseline SFL the dominant term, while $\lambda$ is still large enough for feature evidence to reorder lines where the signals agree.
A smaller value would make the feature signal inert, and a larger one would override the established formula.
As a pre-registered sensitivity analysis, we recompute the primary metric over the fixed grid $\lambda \in \{0.15, 0.25, 0.35, 0.45, 0.55\}$.
This analysis is exploratory and cannot redefine the primary conclusion.

\subsection{Hybrid and Adaptive Ranking Variants}

Beyond the primary random-forest weighting, we report a small, clearly-exploratory family of ranking variants: a hybrid fusion of the available execution signals, a fail-first variant that prioritizes discriminative failing evidence, and an adaptive variant that reweights samples under class imbalance or sparse failing tests.
These variants reuse the same benchmark, mapping, and metric protocol and serve only as interpretation aids.
They are corrected for multiplicity separately and cannot substitute for the primary RF result.
We defer a fully specified learned-fusion mechanism to future work.

\section{Execution Plan}%
\label{sec:execution-plan}

The study runs in four phases: Phases 1--2 produce the rankings for \Cref{rq1}, \Cref{rq2}, and \Cref{rq4}, Phase 3 constructs the ground-truth sets for \Cref{rq3}, and Phase 4 computes metrics and analyses for all research questions.

\subsection{Phase 1: Feature Extraction and Model Training}

For each \TESTS4PY{} subject, we run all tests with \EFDD{} instrumentation using a 120-second timeout per test, convert traces into labeled feature vectors, verify the expected binary or tertiary encoding, train one random forest per subject, and save models and feature vectors so a ranking can be reproduced without rerunning instrumentation.

\subsection{Phase 2: Feature Weighting and Source Line Ranking}

We extract importances, map them to source locations using \Cref{sec:diagnosis-based-fault-localization}, aggregate by maximum importance, normalize weights to $[\varepsilon,1]$, and verify every executable line receives a weight.
For each subject, we compute \TARANTULA{}, \OCHIAI{}, \DSTAR{}, \NAISHTWO{}, and \GPTHIRTEEN{} scores, generate feature-augmented rankings, and average metrics over 10,000 randomized tie-breaking simulations seeded from master seed 42, so arbitrary source order does not dominate cases of equal suspiciousness.

\subsection{Phase 3: Expanded Ground-Truth Discovery}

For each subject, we construct reference patch-hunk sets from official patches and test-impact sets using the line-observation rule from \Cref{sec:variables}.
Test-impact discovery is recorded separately from the patch-hunk set, allowing the analysis to report when a method succeeds only on the reference edit, only on broader impact locations, or on both.

\subsection{Phase 4: Metric Calculation and Analysis}

We calculate the metrics in \Cref{sec:variables} and store them by subject, baseline, method, and metric type, alongside trained models and feature vectors, supporting paired tests, project-level analysis, and independent recomputation.

\section{Analysis Plan}%
\label{sec:analysis-plan}

Directional expectations are positive reference-patch top-$k$ gains for \Cref{rq1}, no systematic degradation in line- or function-level effort for \Cref{rq2}, at least comparable test-impact coverage for \Cref{rq3}, and positive direction across most baselines for \Cref{rq4}.
\Cref{tab:analysis-tiers} separates the primary analysis from the secondary and exploratory and maps each to its research question.

\begin{table}[t]
   \centering
   \caption{Analysis tiers.}%
   \label{tab:analysis-tiers}
   \resizebox{.8\columnwidth}{!}{
   \begin{tabular}{llc}
      \toprule
      \textbf{Tier} & \textbf{Analysis / outcome} & \textbf{RQ} \\
      \midrule
      Primary      & Confounder-adjusted mixed-effects model  & \ref{rq1}--\ref{rq4} \\
      Primary      & Reference-patch top-1/5/10 accuracy      & \ref{rq1} \\
      Primary      & Reference-patch \EXAM{}, wasted effort   & \ref{rq2} \\
      \midrule
      Secondary    & Distribution-free paired tests           & \ref{rq1}--\ref{rq3} \\
      Secondary    & Function-level effort                    & \ref{rq2} \\
      Secondary    & Impact-found@$k$, min.\ rank             & \ref{rq3} \\
      Secondary    & Direction across 5 baselines             & \ref{rq4} \\
      \midrule
      Exploratory  & Ranking variants                         & \ref{rq4} \\
      Exploratory  & $\lambda$/aggregation sensitivity        & n/a \\
      Exploratory  & Importance diagnostics                   & n/a \\
      \bottomrule
   \end{tabular}
   }
\end{table}

\subsection{Primary and Exploratory Analyses}

The primary method is random-forest feature weighting, compared against each of the five unmodified SFL formulas.
The primary outcomes are the five reference-patch metrics: top-1, top-5, and top-10 accuracy, \EXAM{} score, and average-case wasted effort, each computed after randomized tie breaking and averaged per subject.
For a given baseline formula, we call the augmented method \emph{superior} if the confounder-adjusted model's method coefficient is an improvement with a non-negligible standardized effect ($|d|\geq0.2$) on at least four of these five metrics and degrades none of them by a non-negligible effect.
We interpret $H_{\mathrm{ACC}}$ as supported only if RF weighting is superior in this sense for the majority of the five baseline formulas.
We interpret $H_{\mathrm{ROB}}$ as supported only if this superiority holds in all five baseline formulas, confirmed by the method$\times$formula interaction and a sign test.
The exploratory ranking variants form a family with separate multiplicity control.
If the primary comparison is negative or mixed, positive variant results are treated as exploratory evidence about alternative aggregation, not as support for the primary claim.

\subsection{\texorpdfstring{Primary Analysis for \Cref{rq1}, \Cref{rq2}, and \Cref{rq3}}{Primary Analysis for RQ1, RQ2, and RQ3}}

The primary inference for every outcome is a single confounder-adjusted (generalized) linear mixed-effects model fit over all subjects, baselines, and paired method conditions.
Its fixed effects are the method indicator (augmented vs.\ baseline, the effect of interest), the SFL formula as a factor, their interaction, and the confounders of \Cref{sec:variables} as covariates; random intercepts for project and for subject-within-project and a per-project random slope for the method effect estimate effect heterogeneity directly rather than averaging it away.
Continuous metrics (\EXAM{}, wasted effort, function-level effort) use a Gaussian model; binary per-subject outcomes (top-$k$, impact-found@$k$) a binomial logit.
We report the method coefficient with 95\% confidence intervals and standardize it for the superiority rule, read $H_{\mathrm{ROB}}$ from the interaction, and use the confounder coefficients and random-slope variance to characterize where and why the effect varies.
We apply Benjamini--Hochberg correction to the method coefficients within the reference-patch top-$k$, effort, and test-impact families, reported alongside improve/tie/degrade counts per RQ.
With 310 subjects per formula, the design has about 80\% power to detect a standardized method effect of $d\approx0.16$, so smaller effects are not treated as strong practical evidence.

To corroborate the model under minimal distributional assumptions, we re-test each contrast with distribution-free paired procedures: paired permutation tests for top-$k$ and impact-found@$k$ proportions with paired mean differences and 95\% bootstrap confidence intervals, and Wilcoxon signed-rank tests with rank-biserial correlations for \EXAM{}, wasted effort, function-weighted \EXAM{}, faulty-function mass, practical effort, and minimum impact rank.
Patch-hunk and test-impact sets are analyzed separately.
Hypothesis support rests on the model and the four-of-five superiority rule rather than on any single corrected $p$-value, and negligible significant effects are treated as weak practical evidence.
Exploratory hybrid/adaptive tests are corrected separately and reported after the primary results.

\subsection{\texorpdfstring{Robustness Analysis for \Cref{rq4}}{Robustness Analysis for RQ4}}

For the primary RF method, we read the per-formula method effects and the method$\times$formula interaction from the model and determine for each baseline formula whether the augmented method is superior under the four-of-five metric rule defined above.
We require this superiority in all formulas for $H_{\mathrm{ROB}}$ and confirm that the positive direction dominates with a sign test.
We report improve/tie/degrade direction counts per formula.
Exploratory variants are summarized with the same direction counts.

\subsection{Practical Feasibility}

We report medians, interquartile ranges, and per-project distributions of per-phase and total runtime, together with timeouts, exclusions, and trace/storage failures, to judge whether the method is plausible for offline debugging.
Subjects affected by feasibility failures are counted explicitly and not silently removed from denominator summaries.

\subsection{Exploratory Analyses}

To interpret why methods improve or degrade, we analyze feature-category contribution, project-level variation, per-baseline sensitivity, constant-feature fraction, importance concentration/entropy/skewness, failing-vector counts, and feature-to-location mapping-conflict rates.
We do not perform manual relabeling of \TESTS4PY{}, so all diagnostics are computed automatically from artifacts.
We also report the correlation matrix among dependent variables so that redundant metrics are not over-interpreted.
The fixed sensitivity checks vary the lower-bound weight $\varepsilon\in\{10^{-4},10^{-3},10^{-2}\}$, the blend parameter $\lambda\in\{0.15,0.25,0.35,0.45,0.55\}$, and the aggregation operator (mean and sum instead of max).
Where at least three passing and three failing vectors exist, repeated seeded and bootstrap refits quantify feature-importance stability and flag subjects whose rankings may reflect noisy per-subject models.

\subsection{Handling of Edge Cases and Missing Data}

Subjects enter paired analysis if at least one passing and one failing feature vector remain, which is the minimum needed to compute pass-versus-fail feature differences.
They are excluded only if the subject cannot be executed, feature extraction fails (e.g., from the 120-second per-test timeout), or one outcome class is absent.
We impose no higher failing-test threshold, as that would select against sparse but realistic subjects.
We accept partial feature loss, constant-feature removal, and trace-size anomalies unless they make the subject unanalyzable.
All exclusions and deviations will be documented.

\section{Outcome-Neutral Checks}%
\label{sec:outcome-neutral-checks}

\subsection{Data Quality Checks}

Before interpreting results, we verify that labeled feature vectors, binary/tertiary encodings, executable-line weights, and patch-hunk/test-impact records are evaluable.
If a check fails, we record the affected subject and either repair the data-processing step or exclude it from paired analysis with an explicit reason, so implementation failures are not interpreted as localization performance.

\subsection{Sanity Checks on Results}

We add a perfect pass/fail feature as a positive control, replace learned importances with random scores as a negative control, retrain 10 randomly selected subjects with the same seed to verify identical metrics, and inspect paired differences for malformed subjects.
The positive control checks that the pipeline propagates known-fault evidence to the ranking, and the negative control (our primary guard against spurious importance) checks that improvement is not merely an artifact of perturbing suspiciousness scores.

\subsection{Replicability Checks}

We use a master seed ($42$) for reproducible derived seeds, document hyperparameters and constants, rerun to verify identical results, and publish all derived seeds, configurations, and results for every phase with the final manuscript.

\section{Threats to Validity}%
\label{sec:threats-to-validity}

\subsection{Internal Validity}

Using \EFDD{} reduces custom-instrumentation risk but may miss feature types outside its scope, and other learners or hyperparameters might behave differently.
We treat feature importances as ranking evidence rather than a causal explanation of the fault, and use diagnostics to expose concentrated or unstable importance mass.
Feature-to-line mapping uses semantic but heuristic rules, so alternative mappings could change rankings, most of all for features spanning multiple locations such as def-use pairs and scalar relations.
We mitigate this by documenting the rules, recording conflicts, evaluating function-level outcomes that are less brittle to mapping differences, and running an exploratory alternative mapping that assigns def-use and scalar-relation importance to the use line only.

\subsection{External Validity}

\TESTS4PY{} and \EFDD{} are Python-specific, so other languages require adapted instrumentation.
Using all 310 subjects maximizes within-benchmark diversity, but curated faults with reproducible tests and mostly isolated fixes may not transfer to flakiness, partial coverage, or multiple faults.
Subjects with few failing executions may provide weak evidence for learning failure-indicative features; we capture this through feature-dropping, weak-signal, and project-stratified analyses that show whether effects are broad or concentrated.
The SFL baselines are established but do not represent all learning- or specification-based localization approaches.

\subsection{Construct Validity}

Top-$k$, \EXAM{}, and wasted effort inherit single-patch bias when used alone, which the separate function-level, test-impact, and diagnostic families mitigate.
Official fixes define only one reference solution, and test-impact sets are dynamic approximations that depend on the executed tests, so we read them as complementary rather than definitive evidence.
Bug difficulty varies with locality and coupling, which paired within-subject comparisons reduce.

\section{Reproducibility and Data Availability}%
\label{sec:data-availability}

We will publish source code, data, trained models, analysis scripts, and documentation under the Apache 2.0 license as a repository and long-term archival artifact (e.g., Zenodo).

\section{Conclusion}%
\label{sec:conclusion}

This pre-registered study tests whether execution-feature weighting improves statistical fault localization on 310 \TESTS4PY{} Python bugs, that is, whether failure-indicative \EFDD{} features transfer from diagnosis to line-level ranking across SFL baselines.
Its expected contribution is evidence on when richer execution signals improve localization beyond line spectra, with reproducible artifacts for future debugging research.

\section*{Acknowledgments}%
\label{sec:acks}

This work is funded by the European Union (ERC S3, 101093186).
Views and opinions expressed are those of the authors only and do not necessarily reflect those of the European Union or the European Research Council.
Neither the European Union nor the granting authority can be held responsible for them.

\bibliographystyle{IEEEtran}
\bibliography{references}

\end{document}